\documentclass[pra,showpacs,twocolumn]{revtex4}

\usepackage{graphicx}
\usepackage{times}
\usepackage{amsmath}
\usepackage{amsfonts}
\usepackage{amssymb}
\usepackage{bbm}

\begin{document}

\title{Mean-Field Analysis of Spinor Bosons in Optical Superlattices}

\author{Andreas Wagner} 
\author{Andreas Nunnenkamp}
\author{Christoph Bruder}
\affiliation{Department of Physics, University of Basel, 
Klingelbergstrasse 82, 4056 Basel, Switzerland}

\date{\today}

\begin{abstract} 
We study the ground-state phase diagram of spinless and spin-1 bosons
in optical superlattices using a Bose-Hubbard Hamiltonian that
includes spin-dependent interactions. We decouple the unit cells of
the superlattice via a mean-field approach and take into account the
dynamics within the unit cell exactly.  The system supports
Mott-insulating as well as superfluid phases. The transitions between
these phases are second-order for spinless bosons and second- or
first-order for spin-1 bosons. Anti-ferromagnetic interactions
energetically penalize high-spin configurations and elongate all Mott
lobes, especially the ones corresponding to an even atom number on
each lattice site. We find that the quadratic Zeeman effect lifts the
degeneracy between different polar superfluid phases leading to
additional metastable phases and first-order phase transitions.
Finally, we show that an energy offset between the two sites of the
unit cell induces a staircase of single-atom tunneling resonances
which surprisingly survives well into the superfluid regime.
\end{abstract}


\maketitle

\section{Introduction}

Ultracold atom experiments offer the unique opportunity to study
quantum many-body effects in an extremely clean and well-controlled
environment. In contrast to most condensed matter systems they are
characterized by the absence of disorder and other imperfections and
are highly controllable. This is why they have been proposed as
quantum simulators~\cite{lewenstein06, bloch08, bloch12}.

One of the most prominent achievements in this direction has been the
observation of the quantum phase transition between a Mott insulating
and a superfluid phases of ultracold atoms in an optical
lattice~\cite{fisher89,Jaksch1998,greiner02}.  In this situation, the
ratio of the tunneling strength between the lattice sites and the
on-site interaction determines the qualitative behavior of a quantum
gas. If the on-site interaction dominates and the filling is
commensurate the quantum gas is in the Mott-insulating phase. On
the contrary, if the tunneling amplitude is sufficiently large the system
becomes superfluid.

For atoms trapped in a magneto-optical trap the spin degree
of freedom is frozen and the atoms become effectively spinless. If,
however, the quantum gas is trapped by purely optical means, the atoms
keep their spin degree of freedom and the order parameter describing
the superfluid phase becomes a spinor. These systems were first
studied in Refs.~\cite{ho98,ohmi98} and recently reviewed in
Ref.~\cite{stamperkurn12}.

Due to their spin-dependent interactions ultracold spinor quantum
gases in optical lattices offer the possibility to model mesoscopic
magnetism~\cite{wagner11}. They are well described
by the Bose-Hubbard model, but the spin-dependent effects alter the
system in a qualitative and quantitative way 
\cite{imambekov03,tsuchiya04,krutitsky04,kimura05,krutitsky05,pai08}. 
The phase boundaries between  superfluid  and a Mott-insulating
phases are shifted, and for certain atom-configurations the phase
transition is no longer second- but first-order. This is a consequence
of the additional spin-dependent on-site interaction. If this
interaction is anti-ferromagnetic, atomic singlets are energetically
favored and the Mott-insulating phase is stable in some parameter
ranges where the system is superfluid for spinless atoms. The
occurrence of first-order phase transitions enables metastable phases and the system shows hysteretic behavior.

\begin{figure}[tb]
\begin{center}
\includegraphics[width=0.4 \textwidth]{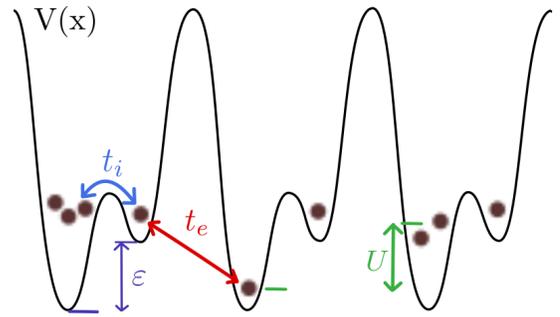}
\caption{(Color online) Potential landscape of an optical superlattice and parameters of the Bose-Hubbard model
  (\ref{spinless-mfham}) for spinless ultracold atoms in optical
  superlattices. The dots depict the atoms in the superlattice
  potential $V(x)$. $\varepsilon$ is the energy offset between the two
  sides of the double well, $t_i$ ($t_e$) is the intra- (inter-)well
  tunneling amplitude, and $U$ is the strength of the on-site
  interaction.}
\label{fig1:superlattice}
\end{center}
\end{figure}

In this paper we investigate ultracold bosons loaded into optical
superlattices. These systems have already been studied theoretically
\cite{sebby06,rey07} and experimentally
\cite{anderlini07,folling07,lee07,trotzky08}. We take into account the
dynamics in the double wells exactly and include the tunneling between
neighboring unit cells via a mean-field ansatz. The system supports
Mott-insulating phases as well as  superfluid
phases~\cite{buonsante05,chen10}. The former are characterized by a
fixed number of atoms per unit cell. For spin-1 atoms we find that
some of the phase transitions become first order similar to the case
of usual period-1 lattices~\cite{krutitsky05}.  

We include the effects of magnetic fields by using an effective Hamiltonian which includes a quadratic Zeeman shift. For anti-ferromagnetic interactions magnetic fields break the degeneracy between different polar superfluid phases. This leads to new classes of
metastable phases and changes the phase boundaries significantly. In the
ferromagnetic case magnetic fields cause first-order phase transitions
and metastable phases. These results apply to spin-1 atoms in
superlattices as well as in usual lattices.

Finally, we examine single-atom
tunneling resonances in superlattices.  These are known from isolated
double-well potentials~\cite{averin08,cheinet08,wagner11} and are
generalized to superlattices in this work.

The remainder of this paper is organized as follows. In
Sec.~\ref{sec:spinless} we introduce the mean-field Hamiltonian for
spinless bosons in optical superlattices. We discuss methods to treat
this Hamiltonian and present the phase diagram in
Sec.~\ref{sec:spinlessphasediagram}. Section~\ref{sec:sat} treats the
phenomenon of single-atom resonances for spinless atoms in optical
superlattices.  In Sec.~\ref{sec:spin} we generalize the
Bose-Hubbard Hamiltonian of Sec.~\ref{sec:spinless} by including
spin-dependent interactions. In Sec.~\ref{sec:spin1phasediagram} the
phase diagram of spin-1 atoms is examined. We discuss the novel
aspects due to the spinor nature of the order parameter and point out
differences in the single-atom resonances. Finally, we include
magnetic fields in Sec.~\ref{sec:mag} which enhance spin-dependent
effects and lead to additional metastable phases.

\section{Spinless Bosons}
\label{sec:spinless}

\begin{figure}[tb]
\begin{center}
\includegraphics[scale=.75,trim=1.8cm 3cm 0cm 2cm]{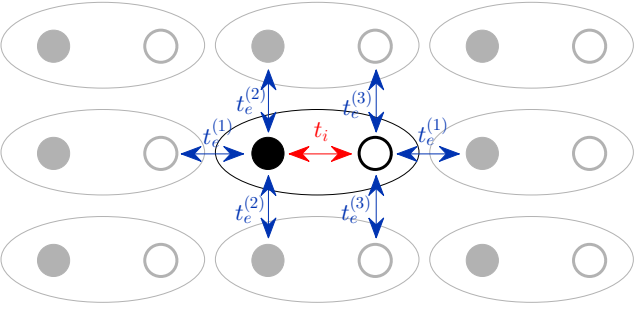}
\caption{(Color online) Sketch of the two-dimensional period-2
  superlattice. The filled circles depict the left wells of each unit
  cell, the open ones the right wells. The tunneling amplitude within
  the unit cells is $t_i$. There are three different tunneling
  processes between neighboring unit cells, $t_e^{(1)}$, $t_e^{(2)}$,
  and $t_e^{(3)}$.}
\label{fig2:array} 
\end{center}
\end{figure}

Optical lattices are arrays of weakly connected optical micro-traps
created by interfering counter-propagating laser beams. When the
lattice is superimposed with a second standing laser beam with half
the wavelength an optical superlattice is created (see
Fig.~\ref{fig1:superlattice}). The unit cell of a period-2
superlattice is a double-well potential~\cite{anderlini07,folling07,lee07,trotzky08}.

Ultracold bosons in sufficiently deep optical lattices can be
described by the Bose-Hubbard
Hamiltonian~\cite{Jaksch1998,vaucher08}. Here we examine the physics
of bosonic atoms in optical superlattices when the overall atom
density is chosen such that there are a few atoms per double well. If
one neglects tunneling between neighboring unit cells the atoms in
each unit cell can be described by the Hamiltonian
\begin{eqnarray}
\label{spinless-singleham}
&\hat H_0&=\frac{U}{2} \sum_{k=L,R} \hat n_k(\hat n_k-1)-t_i (\hat L^\dagger \hat R 
+h.c.)  \nonumber \\
&+&\varepsilon \left( \hat n_L-\hat n_R \right) - \mu \left( \hat n_L+\hat n_R \right),
\end{eqnarray}
where $\hat L$ $(\hat L^\dagger)$ and $\hat R$ $(\hat R^\dagger)$ are
bosonic annihilation (creation) operators for atoms in the left or
right well, $\hat n_L$ $(\hat n_R)$ is the atom number operator at the
left (right) site. $U$ is the on-site interaction and $t_i$ is the
tunneling strength between the sites of the double well. The energy
offset between the sites is given by $\varepsilon$ and the chemical
potential is $\mu$ (see Fig.~\ref{fig1:superlattice}).  The parameters
can be tuned by changing the intensity and the phase difference
between the counter-propagating laser beams; it is possible to tune the system
from the regime of strong tunneling ($t_i \gg U$) to the regime of weak
tunneling ($t_i \ll U$).

The Hamiltonian of an array of connected double-well potentials
includes tunneling between neighboring unit cells. We choose the
configuration as shown in Fig.~\ref{fig2:array} where there are in
general three different inter-well tunneling amplitudes, $t_e^{(1)}$,
$t_e^{(2)}$ and $t_e^{(3)}$. It turns out that the our results depend
only weakly on the differences among the inter-well tunneling
amplitudes. This is why we will assume $t_e = t_e^{(1)} = t_e^{(2)} =
t_e^{(3)}$.

Within the mean-field approximation which has been developed in 
Refs.~\cite{sheshadri93, buonsante05}, our Hamiltonian reads
\begin{eqnarray}
\label{spinless-mfham}
&\hat H &=\hat H_0  - t_e\Big( \phi_R \hat L^\dagger +\phi_L \hat R^\dagger +2 z \phi_R \hat R^\dagger  +2 z \phi_L \hat L^\dagger  \nonumber \\
&-& \phi_R \phi_L^*  -  z \phi_R \phi_R^* -  z\phi_L^* \phi_L + h.c. \Big)
\end{eqnarray}
where we introduced the mean-field parameters $\phi_R=\langle \hat
R\rangle $, $\phi_L=\langle \hat L\rangle$, and $z=1$ for 2D lattices
and $z=2$ for 3D lattices.  The Hamiltonian treats the internal
degrees of freedom of each unit cell exactly and approximates the
tunneling between the unit cells via a mean-field ansatz.  This
approach is expected to give satisfactory results if the tunneling
inside the unit cells is stronger than the tunneling between the unit
cells (i.e., $t_e<t_i$), otherwise correlations between neighboring
double wells would be stronger than correlations within the double
wells and should not be neglected.

\begin{figure}[tb]
\begin{center}
\includegraphics[scale=.62,trim=1cm 0cm 0cm 0cm]{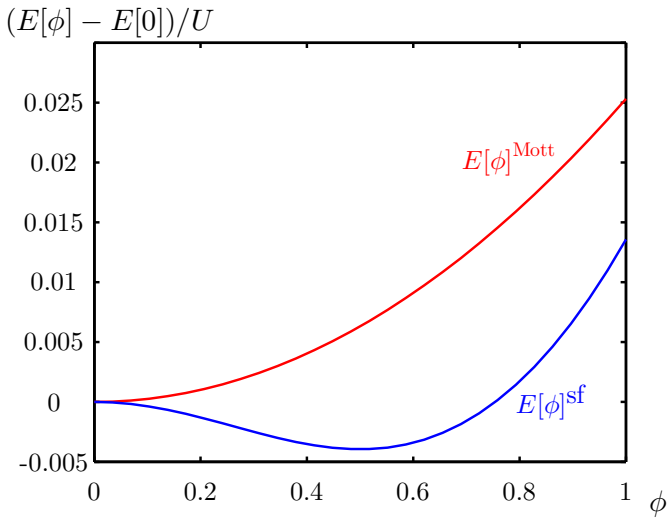}
\caption{ (Color online) Ground-state energy $E[\phi_L,\phi_R]$ of the
  Hamiltonian (\ref{spinless-mfham}) as a function of the order
  parameter $\vec \phi=\{\phi_L,\phi_R \}$. For a symmetric unit cell
  $\varepsilon=0$, we have $\phi_L=\phi_R=\phi$. The red line
  $E[\phi]^\text{Mott}$ corresponds to the Mott-insulating phase
  ($\mu/U=0.2$, $t_i/U=0.05$, and $t_e/U=0.005$), the blue line
  $E[\phi]^\text{sf}$ to the superfluid phase ($\mu/U=0.5$,
  $t_i/U=0.22$, and $t_e/U=0.022$).  }
\label{fig3:spinless-super-efunctionals}
\end{center}
\end{figure}

The system is in the Mott-insulating phase if $\phi_L = \phi_R = 0$
and in the superfluid phase if $\phi_L \not= 0 \not= \phi_R$. In the
latter case the number of superfluid atoms $n_L^{\text{sf}}$ and
$n_R^{\text{sf}}$ on the left and right site is given by $\vec \phi =
(\phi_L, \phi_R) = (\sqrt{n_L^{\text{sf}}}, \sqrt{n_R^{\text{sf}}})$.

There are several possibilities to treat the Hamiltonian
(\ref{spinless-mfham}).  For a given set of parameters $\{\mu,
\varepsilon, t_i,t_e,U \}$ the task is to find the self-consistent
values of $\phi_L=\langle \psi_{\vec \phi}^{(0)}| \hat L | \psi_{\vec
  \phi}^{(0)} \rangle$, and $\phi_R=\langle \psi_{\vec \phi}^{(0)}|
\hat R | \psi_{\vec \phi}^{(0)} \rangle$ where $|\psi_{\vec
  \phi}^{(0)}\rangle$ denotes the ground state of the Hamiltonian
(\ref{spinless-mfham}) for a given order parameter $\vec \phi$.  On
the one hand the self-consistent values are fixed points of the map
\begin{eqnarray}
\label{spinlessmap}
\vec \phi_{i+1}=\{\langle \hat R \rangle_{\vec \phi_i},\langle \hat L\rangle _{\vec \phi_i}\}
\end{eqnarray}
where the index $i$ refers to the $i$th step in the iterative
procedure used to find the self-consistent value of the order
parameter. On the other hand the self-consistent values of $\vec \phi$
correspond to the local extrema of the energy functional
\begin{eqnarray}
E[\phi_L,\phi_R]=\langle\psi_{\vec \phi}^{(0)}|  \hat{H}| \psi_{\vec \phi}^{(0)} \rangle
\end{eqnarray}
and its local minima correspond to stable fixed points of the
map~(\ref{spinlessmap}) which can be found by the iterative procedure.

\begin{figure}[tb]
\begin{center}
\includegraphics[width=0.49\textwidth]{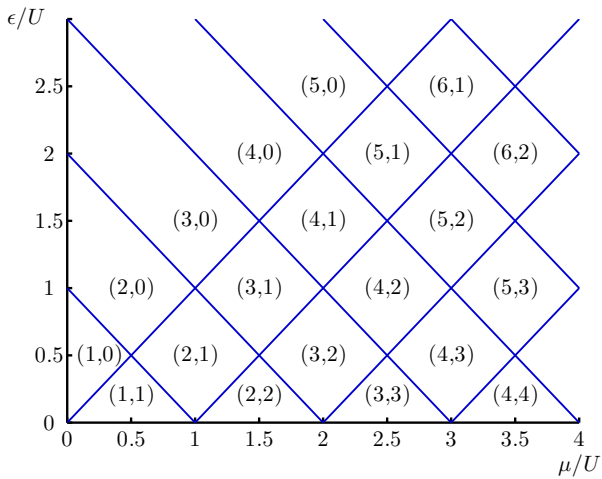}
\caption{Phase diagram of spinless bosons in a superlattice described
  by the Hamiltonian (\ref{spinless-mfham}).
  In the atomic limit $t_i=t_e=0$
  the Hamiltonian is diagonal in the Fock basis
  and supports only Mott-insulating phases. The blue lines mark the
  phase boundaries and $(n_L,n_R)$ denotes the occupation of the left
  and the right site in the double well.}
\label{fig4:spinless-mottphases} 
\end{center}
\end{figure}

\begin{figure}[tb]
\begin{center}
\includegraphics[width=0.5\textwidth]{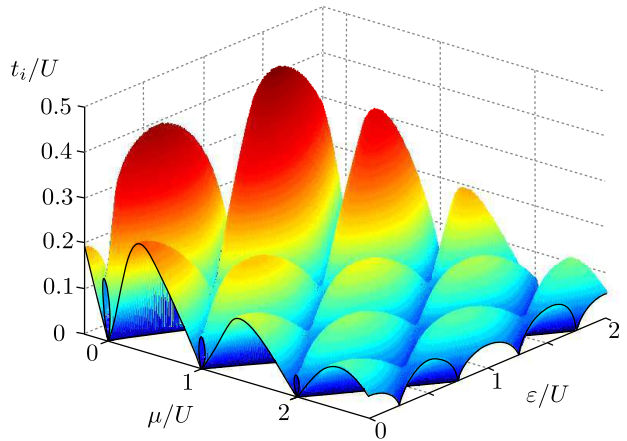}
\caption{ (Color online) Phase diagram of spinless bosons in a
  two-dimensional superlattice described by the Hamiltonian
  (\ref{spinless-mfham}).  We plot the critical internal tunneling
  amplitude $t_i$ as a function of chemical potential $\mu$ and energy
  offset $\varepsilon$ for $t_i=10 t_e$.  In
  Fig.~\ref{fig4:spinless-mottphases} we show a cut through this 3D
  plot at $t_i=t_e=0$ and in Fig.~\ref{fig6:spinless-mott2} at
  $t_i/U=0.05$. The edge at $\varepsilon=0$ of the phase diagram
  reveals the contraction of Mott lobes to loops at integer values of
  $\mu/U$.}
\label{fig5:spinless-ratio10}
\end{center}
\end{figure} 

\begin{figure}[tb]
\begin{center}
\includegraphics[width=0.49\textwidth]{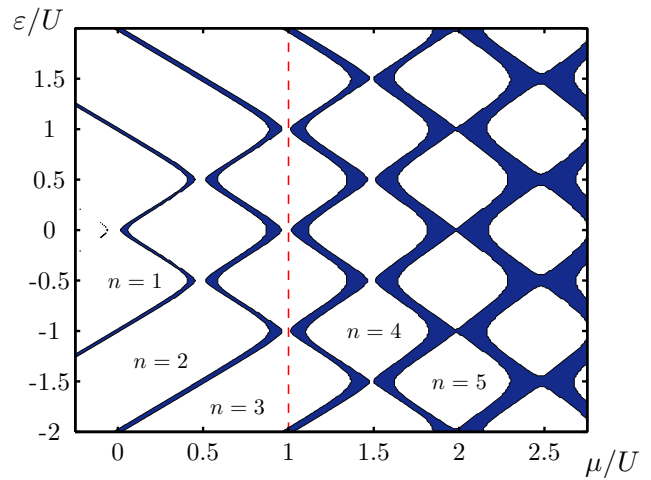}
\caption{ (Color online) Phase diagram of spinless bosons in a
  two-dimensional superlattice described by the Hamiltonian
  (\ref{spinless-mfham}) for $t_i/U=10t_e/U=0.05$. The
  blue areas depict the superfluid phase. The dashed line corresponds to the
  parameters chosen for Fig.~\ref{fig7:spinless-treppe1}.
 }
\label{fig6:spinless-mott2}
\end{center}
\end{figure}

For the Hamiltonian (\ref{spinless-mfham}) there are only two classes
of energy functionals (see
Fig.~\ref{fig3:spinless-super-efunctionals}): those with only one
local extremum at $\phi_L=\phi_R=0$ corresponding to a Mott-insulating
phase, and those with a second extremum at $\phi_L \not= 0 \not=
\phi_R$ which is the global minimum and corresponding to a superfluid
phase.

This enables us to distinguish the Mott and superfluid quantum phases
with minimal numerical effort. We only have to calculate the
ground-state energy for $\vec \phi=0$ and in its proximity $\vec \phi
\approx 0$. If $E[\vec \phi \approx 0]-E[\vec\phi=0]$ is positive, the
system is Mott-insulating; if it is negative the system is superfluid.

\subsection{The phase diagram for spinless bosons}
\label{sec:spinlessphasediagram}

In this section we determine the ground-state phase diagram of the
Hamiltonian (\ref{spinless-mfham}). For a chemical potential $\mu$, an
energy offset $\varepsilon$, and a given ratio of the tunneling
amplitudes $t_e/t_i$ we calculate the critical tunneling amplitude
$t_e$ above which the system is superfluid. In the following we use
the on-site interaction $U$ as the unit of energy.

In Fig.~\ref{fig4:spinless-mottphases} we plot the ground state as a
function of chemical potential $\mu$ and offset $\varepsilon$ in the
atomic limit $t_i=t_e=0$. In this case, the Hamiltonian
(\ref{spinless-mfham}) is diagonal in the Fock basis and the system
supports only Mott phases $(n_L,n_R)$ characterized by the number of
atoms in the left $n_L$ and right well $n_R$. Each of the diamonds in
Fig.~\ref{fig4:spinless-mottphases} corresponds to one Fock state,
i.e., a fixed particle number in the unit cell as well as a fixed
particle number in the left and the right site of each unit cell. When
we increase $\mu/U$ for fixed energy offset $\varepsilon/U$ the number
of particles in the unit cells increases while the ratio between left
and right remains similar. When we increase $\varepsilon/U$ for fixed
chemical potential $\mu/U$ the atom number is constant but the atom
distribution within the unit cells changes. Because we set the
tunneling to zero this happens non-continuously. 
Figure~\ref{fig4:spinless-mottphases} is mirror-symmetric along the
$\varepsilon=0$ axis, i.e., when $\varepsilon \rightarrow -
\varepsilon$ the atom number distribution is inverted,
$(n_L,n_R)\rightarrow(n_R,n_L)$.

In Fig.~\ref{fig5:spinless-ratio10} we plot the critical tunneling
strength $t_i$ at which the system becomes superfluid as a function of
the chemical potential $\mu$ and the offset $\varepsilon$ for
$t_i/t_e=10$. It is convenient to pick a fixed ratio of $t_i/t_e$ in
order to obey the constraint $t_e<t_i$. For fixed energy offset
$\varepsilon$ we recover Mott lobes, which are familiar from the case
of a usual lattice~\cite{fisher89}, although the Mott phase for atoms
in superlattices is characterized by a fixed integer atom number per
unit cell, i.e. $n=\langle \hat n_L +\hat n_R \rangle$ where $n$ is an
integer number. When $\varepsilon/U$ has an integer value the Mott lobes
corresponding to an odd atom number per unit cell contract to Mott loops
and if $\varepsilon/U$ has an half-integer value the lobes corresponding
to an even atom number contract to loops~\cite{buonsante05}. As the
energy offset $\varepsilon$ is varied, the size of the Mott lobes changes
and they constitute tubes of fixed integer atom number per unit cell.
The base of the plot (i.e., the $t_i=t_e=0$ plane) shows the diamond
structure given in Fig.~\ref{fig4:spinless-mottphases}. The nodes of the
diamonds are special: these are the values of the energy offset
$\varepsilon$ where the lobes contract to loops, i.e. the Mott tubes
touch the $t_i=t_e=0$ plane only at one point.

Figure~\ref{fig6:spinless-mott2} presents a cut through
Fig.~\ref{fig5:spinless-ratio10} at $t_i/U=0.05$, showing the Mott
insulating phases in white and the superfluid phases in blue.  The
Mott diamonds of Fig.~\ref{fig4:spinless-mottphases} are connected for
non-vanishing tunneling amplitudes. This means that the quantum
numbers $(n_L, n_R)$ change continuously when $\varepsilon$ is varied
and the Mott insulating phases are characterized by one number $n=n_L
+ n_R$, the total number of particles per unit cell. In
Fig.~\ref{fig6:spinless-mott2} the chosen tunneling amplitudes are too
large to see the connections between the Mott diamonds for $n \geq 5$,
nevertheless the quantum numbers $(n_L, n_R)$ are not fixed to integer
values for these Mott phases either.

\begin{figure}[tb]
\begin{center}
\includegraphics[width=0.5\textwidth]{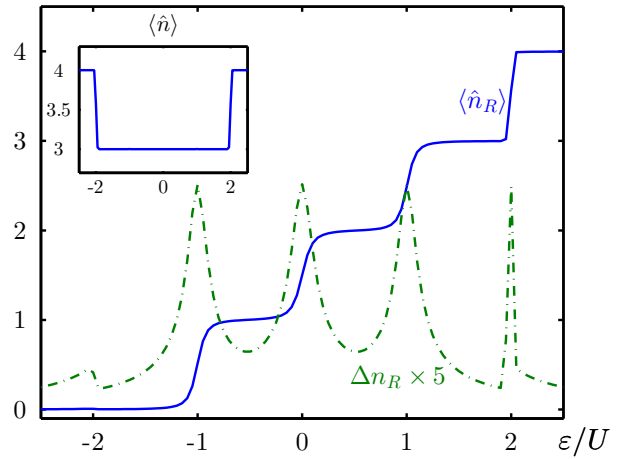}
\caption{ (Color online) Mean particle number in the right well
  $\langle \hat n_R \rangle$ (solid) and standard deviation $\Delta
  n_R=\sqrt{\langle \hat{n}_R^2 \rangle - \langle \hat{n}_R
    \rangle^2}$ (dashed) as a function of the energy offset
  $\varepsilon$ for Hamiltonian (\ref{spinless-mfham}) with $\mu/U=1$
  and $t_i/U=10t_e/U=0.05$.  The inset shows the mean total particle
  number in the double well $\langle \hat n \rangle=\langle \hat n_L+
  \hat n_R \rangle$.  The standard deviation is multiplied by 5.  }
\label{fig7:spinless-treppe1}
\end{center}
\end{figure}

\begin{figure}[tb]
\begin{center}
\includegraphics[width=0.5\textwidth]{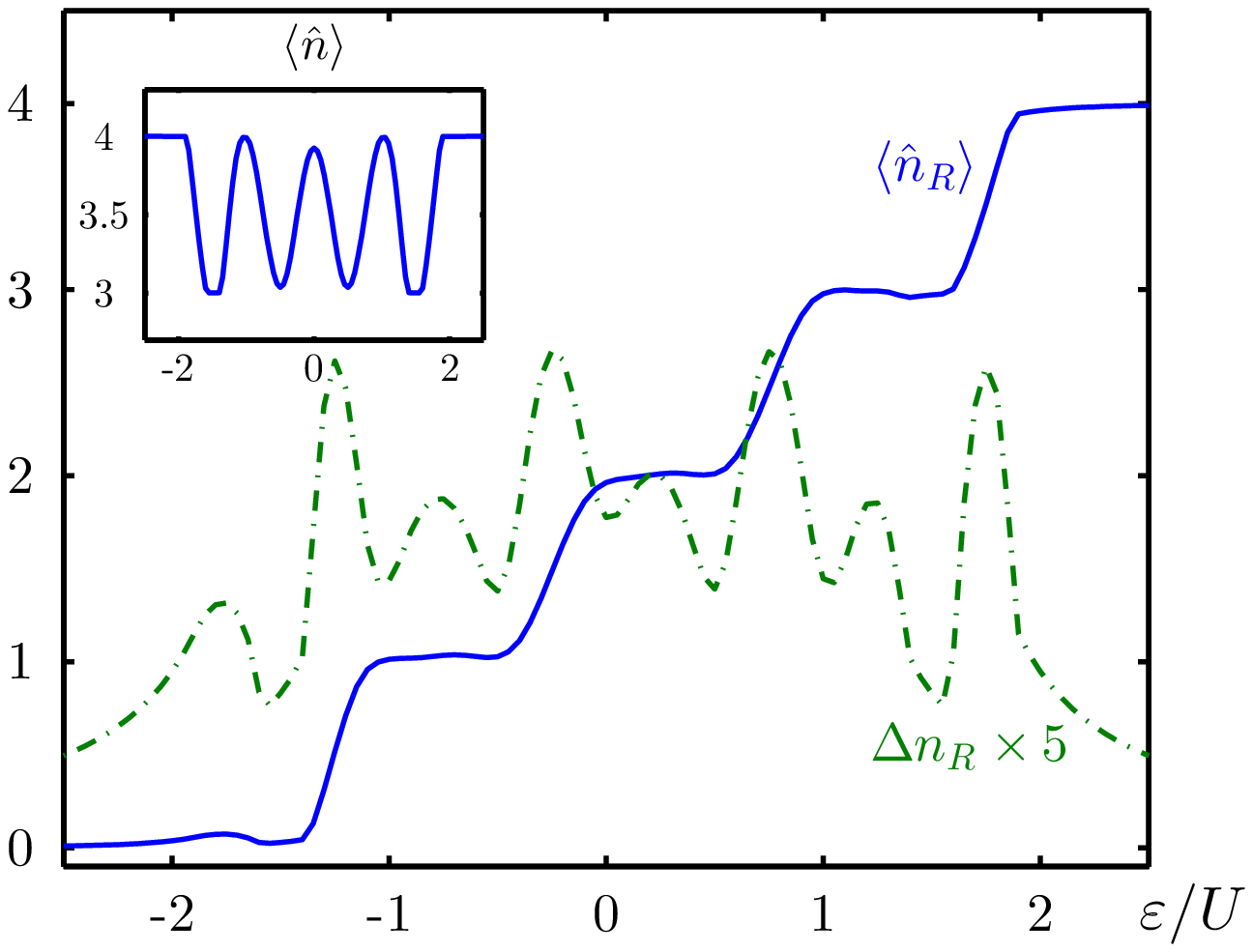}
\caption{ (Color online) Mean particle number in the right well
  $\langle \hat n_R \rangle$ (solid) and standard deviation $\Delta
  n_R= \sqrt{\langle \hat{n}_R^2 \rangle - \langle \hat{n}_R
    \rangle^2}$ (dashed) as a function of the energy offset
  $\varepsilon$ for Hamiltonian (\ref{spinless-mfham}) with
  $\mu/U=1.2$ and $t_i/U=10t_e/U=0.1$.  The inset shows the total
  particle number in the double well $ \langle \hat n \rangle =\langle
  \hat{n}_L + \hat{n}_R \rangle$.  The standard deviation is
  multiplied by 5.  }
\label{fig8:spinless-treppe3}
\end{center}
\end{figure} 

\begin{figure}[tb]
\begin{center}
\includegraphics[width=0.49\textwidth]{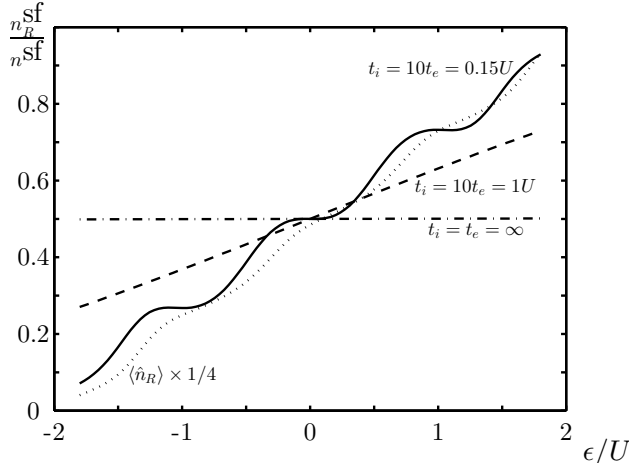}
\caption{Fraction of condensed particles in the right well
  $n_R^{\text{sf}}/n^{\text{sf}}$ (\ref{nlsf}) for
  spinless atoms described by the Hamiltonian (\ref{spinless-mfham}). The
  parameters are chosen such that the system is superfluid over the
  full range of $\varepsilon/U \in [-1.8,1.8]$. The chemical potential
  for all curves is $\mu/U = 1.25$. The solid line corresponds to
  $t_i/U = 10 t_e/U = 0.15$, the dashed line to $t_i/U = 10 t_e/U = 1$
  and the dashed-dotted line to $t_i = t_e = \infty$. The dotted line
  shows the atom number in the right well $\langle \hat n_R\rangle$ as
  a function of $\varepsilon$ for $t_i/U = 10 t_e/U = 0.15$.}
\label{fig9:spinless-sf-treppe}
\end{center}
\end{figure}

\subsection{Single-atom resonances within superlattices}
\label{sec:sat}

If the offset $\varepsilon$ is changed with the other parameters
$\mu$, $t_i$, and $t_e$ fixed, the atom number distribution within the
double wells becomes asymmetric. Due to the finite on-site interaction
the atom numbers do not change proportionally to $\varepsilon$ but in
steps, which are called single-atom resonances.  For isolated
double-well potentials these bosonic staircases were theoretically
predicted \cite{averin08,rinck11,wagner11} and experimentally detected
\cite{cheinet08}.

Within the Mott-insulating phase the superlattice decomposes into an
array of isolated double wells. Thus, it is possible in this regime to
observe the same bosonic staircases as in the case of single
double-well potentials.

Figure~\ref{fig7:spinless-treppe1} shows the mean atom number in the
right well $\langle \hat n_R \rangle$ and the standard deviation
$\sqrt{\langle \hat n_R^2 \rangle - \langle \hat n_R \rangle^2}$ along
the dashed line shown in Fig.~\ref{fig6:spinless-mott2}. The system is
Mott-insulating for a large parameter range ($\mu/U = 1$,
$t_i/U=0.05$, $t_e/U=0.005$, and $-1.9 \le \varepsilon/U \le 1.9$).
The inset shows the total number of atoms in the double well $\langle
\hat{n} \rangle = \langle \hat{n}_L + \hat{n}_R \rangle $ which
changes discontinuously at $\varepsilon/U \approx \pm 2$ signaling
that the system leaves the Mott-insulating phase. This causes a
discontinuity in the function $\langle \hat{n}_R\rangle$. The steps
are equidistant since the difference in the on-site interaction
between one and two atoms and between two and three atoms (and for
higher atom numbers) are the same, namely equal to the on-site
interaction $U$. The spacing between two steps is $\Delta
\varepsilon=U$, i.e., the steps occur when the energy offset is large
enough to compensate the on-site interaction.

In Fig.~\ref{fig8:spinless-treppe3} the expectation value $\langle
\hat n_R\rangle$ is plotted along a curve with stronger tunneling
amplitudes and a shifted chemical potential as compared to
Fig.~\ref{fig7:spinless-treppe1}. Along this curve the system is
mostly superfluid. This can be seen from the inset of
Fig.~\ref{fig8:spinless-treppe3}. For $-2U \le \varepsilon \le 2 U$ the
particle number per double well $\langle \hat{n} \rangle$ is not constant anymore.  Although the atom staircase looks similar to the one in
Fig.~\ref{fig7:spinless-treppe1} there are significant differences.
The fluctuations in the particle number $\Delta n_R$ are greatly
increased and additional maxima appear. These number fluctuations can
be measured in experiments~\cite{cheinet08}.

Single-atom resonances can also be seen in the density of condensed
atoms. The number of condensed atoms per site is connected to the value
of the order parameter via the relations, $n_L^{\text{sf}}=|\langle \hat
L \rangle|^2$ and $n_R^{\text{sf}}=|\langle \hat R \rangle|^2$.
The ratio of condensed atoms in the right well to the number of condensed
atoms in both wells,
\begin{eqnarray}
\label{nlsf}
\frac{n_R^{\text{sf}}}{n^{\text{sf}}}=\frac{|\langle \hat R
  \rangle|^2}{|\langle \hat L \rangle|^2+|\langle \hat R \rangle|^2}\:,
\end{eqnarray}
is plotted in Fig.~\ref{fig9:spinless-sf-treppe} along several
paths in parameter space. The solid line corresponds to $t_i=10
t_e=0.15 U$ and shows steplike behavior. The dotted line shows the
total atom number in the right well $\langle \hat n_R\rangle$ for the same
tunneling amplitudes. Note that the steplike structure is more pronounced
for the density of condensed atoms than for the total atom number.
For higher tunneling rates the staircase structure disappears (see dashed
line with $t_i=10 t_e=1 U$). In the limit of infinite tunneling amplitudes
the ratio $n_{L/R}^{\text{sf}}/n^{\text{sf}}$ does not depend on $\varepsilon$
because there is an infinite amount of atoms in the unit cell
and changing $\varepsilon/U$ by one moves only one atom from one site to
the other. This analysis of the asymmetry of the superfluid density
also helps to understand the connection between $\phi_L$ and $\phi_R$.
The ratio of the two order parameters $\phi_L$ and $\phi_R$ does not
only depend on the energy offset $\varepsilon$ but also on the
tunneling amplitudes, the chemical potential, and the on-site
interaction.

\section{Spinor Bosons}
\label{sec:spin}

The Bose-Hubbard model of Eq.~(\ref{spinless-mfham}) can be
generalized to spin-1 atoms by taking a spin-dependent on-site
interaction into account \cite{imambekov03}.
To obtain a double-well mean-field Hamiltonian for spin-1 atoms in
optical superlattices we have to replace in Eq.~(\ref{spinless-mfham})
the annihilation and creation operators, as well as the order
parameters, by 3-component vectors according to the three hyperfine
projections for spin-1 bosons and add a term containing the
spin-dependent on-site interaction. The resulting Hamiltonian is
\begin{eqnarray}
\label{spin-mfham}
&\hat{H}&=\frac{U_0}{2} \sum_{k=L,R} \hat n_k(\hat n_k-1)-t_i 
\left(\hat{\textbf{L}}^\dagger \cdot \hat{\textbf{R}} +h.c.\right)  \nonumber \\ 
&+&\varepsilon \left( \hat n_L-\hat n_R \right)  - \mu \left( \hat n_L+\hat n_R \right) \nonumber \\
&+&    
\frac{U_2}{2} \sum_{k=L,R}\left(  \hat{\textbf{S}}_k^2 -2\hat n_k\right)  \nonumber \\
&-&t_e  \Big[ \vec\phi_{R}\cdot  \hat{ \textbf{L}}^\dagger   +\vec\phi_{L} \cdot  \hat {\textbf{R}}^\dagger  +2 z \vec\phi_{L} \cdot  \hat {\textbf{L}}^\dagger +2 z \vec\phi_{R} \cdot  \hat {\textbf{R}}^\dagger  \nonumber \\
&-&  \vec\phi_{ R}  \cdot \vec\phi_{ L}^*  -  z\vec\phi_{ L}  \cdot \vec\phi_{ L}^* -  z  \vec\phi_{ R}^*  \cdot \vec\phi_{ R}
+ h.c.\Big]
\end{eqnarray}
where $\hat{\textbf{L}}^\dagger =\{\hat L_1^\dagger,\hat
L_0^\dagger,\hat L_{-1}^\dagger\} $ is a vector containing the
creation operators for the left well, i.e., $\hat L_m$ creates an atom
in the $m$th hyperfine state in the left well. Similarly, $\hat
{\textbf{L}}$ consists of annihilation operators of the left well and
$\hat{\textbf{R}} \ (\hat{\textbf{R}}^\dagger)$ are annihilation
(creation) operators for the right well. $\hat n_L=\sum_\sigma \hat
L^\dagger_\sigma \hat L_\sigma$ $\left(\hat n_R=\sum_\sigma \hat
R^\dagger_\sigma \hat R_\sigma \right)$ is the atom number at the left
(right) site.  The annihilation and creation operators obey the
canonical commutation relations $[\hat L_i,\hat L^\dagger_j]=[\hat
  R_i,\hat R^\dagger_j]=\delta_{ij}$ and $[\hat R_i,\hat
  L^\dagger_j]=[\hat L_i,\hat
  R^\dagger_j]=0$. $\hat{\textbf{S}}_L=\sum_{\sigma \sigma'} \hat
L^\dagger_\sigma \hat{\textbf{T}}_{\sigma \sigma'} \hat L_{\sigma'}$
is the total spin on the left site and the total spin on the right
site is $\hat{\textbf{S}}_R=\sum_{\sigma \sigma'} \hat
R^\dagger_\sigma \hat{\textbf{T}}_{\sigma \sigma'} \hat R_{\sigma'}$,
where $\hat{\textbf{T}}_{\sigma \sigma'}$ are the usual spin-1
matrices. The dimensionality of the array is contained in the
parameter $z$; for 2D lattices $z=1$ and for 3D lattices $z=2$. The
vectors $\vec\phi_{ L}=\{\phi_{L}^{(1)}, \phi_{L}^{(0)},
\phi_{L}^{(-1)}\}$ and $\vec\phi_{ R}=\{\phi_{R}^{(1)},
\phi_{R}^{(0)}, \phi_{R}^{(-1)}\}$ contain the six order parameters of
the Hamiltonian~(\ref{spin-mfham}). Note that the system is
rotationally symmetric and $\phi_1=\phi_{-1}$ 
for both $\vec \phi_L$ and $\vec \phi_R$~\cite{ho98}.

The term proportional to $U_2$ describes spin-dependent contact
interactions: in the case of anti-ferromagnetic interactions $U_2>0$
(e.g.~$^{23}$Na) it penalizes non-zero spin configurations while it favors
high-spin configurations in the case of ferromagnetic interactions
$U_2<0$ (e.g.~$^{87}$Rb). Whereas the ratio $t/U_0$ can be controlled
with the intensity of the laser beams, the ratio $U_2/U_0$ depends on
the spin-2 and spin-0 scattering lengths of the spin-1
atoms~\cite{ho98}, e.g.~$U_2/U_0=0.04$ for $^{23}$Na.

\begin{figure}[tb]
\begin{center}
\includegraphics[width=0.49\textwidth]{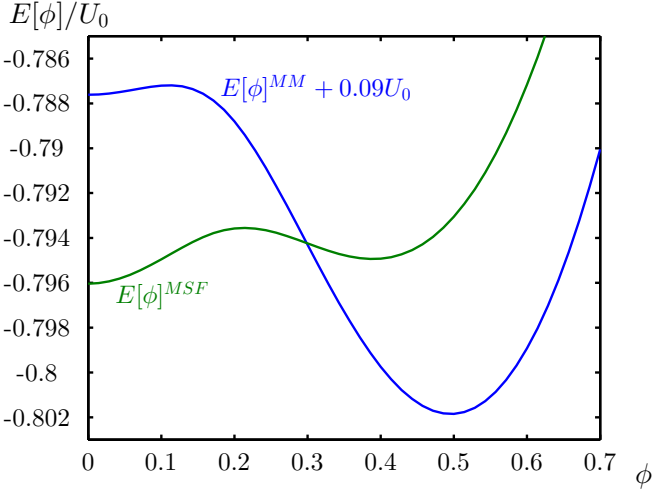}
\caption{(Color online) Ground-state energy $E[\vec
    \phi_L,\vec\phi_R]$ of the Hamiltonian (\ref{spin-mfham}) as a
  function of the order parameter $\vec \phi$. The unit cell is chosen
  to be symmetric $\varepsilon=0$, i.e., we have
  $\vec\phi_L=\vec\phi_R=\vec\phi$, we have chosen $\phi_0=0$, and 
  $\phi_1=\phi_{-1}$ due to symmetry constraints. 
  The blue line $E[\phi]^{MM}$ corresponds a point in parameter space
  ($\mu/U_0=0.25$, $t_i/U_0=0.35$, $t_e/U_0=0.035$, $U_2/U_0=0.04$ ,
  and $\varepsilon=0$) where there is a metastable Mott phase in
  addition to the stable superfluid phase.  The green line
  $E[\phi]^{MSF}$ corresponds a point within the metastable superfluid
  phase ($\mu/U_0=0.25$, $t_i/U_0=0.3$, $t_e/U_0=0.03$,
  $U_2/U_0=0.04$, and $\varepsilon=0$).  $E[\phi]^{MM}$is shifted by
  $0.09U_0$ to show the two curves in the same plot.  }
\label{fig10:spin1-super-efunctionals}
\end{center}
\end{figure}

\begin{figure}[tb]
\begin{center}
 \includegraphics[width=0.49\textwidth]{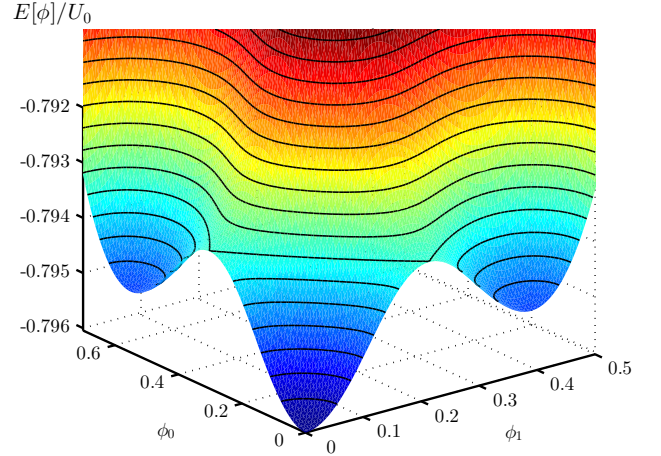}
\caption{ (Color online) Energy functional $E[\vec \phi_L,\vec\phi_R]$
  for the Hamiltonian (\ref{spin-mfham}) with anti-ferromagnetic
  interactions ($t_i/U_0=0.3$, $t_e/U_0=0.03$, $U_2/U_0 = 0.04$,
  $\mu/U_0 = 0.25$, $\varepsilon = 0$). Because the unit cell is
  chosen to be symmetric the order parameters of the left and the
  right are equal ($\vec\phi=\vec\phi_L = \vec\phi_R$) and the
  corresponding indices are suppressed. Because of the rotational
  symmetry $\phi_1=\phi_{-1}$.
 }
\label{fig11:spin1-efsurf}
\end{center}
\end{figure}

The Hamiltonian (\ref{spin-mfham}) has a much richer phase diagram
than the Hamiltonian (\ref{spinless-mfham}).  In addition to
Mott-insulating and superfluid quantum phases, the spin-1 Bose-Hubbard
model gives rise to metastable quantum phases. This can be seen by
looking at the energy functionals $E[\vec \phi_L,\vec \phi_R]$ of the
Hamiltonian (\ref{spin-mfham}). In addition to those shown in
Fig.~\ref{fig3:spinless-super-efunctionals}, two other classes of
energy functionals arise for anti-ferromagnetic spin interactions
$U_2>0$, see Fig.~\ref{fig10:spin1-super-efunctionals}.  These energy
functionals have two local minima and an iterative procedure similar
to the one described in Sec.~\ref{sec:spinless} does not lead to an
unique value of the order parameter but depends on the starting
point. When the starting point of the iterative procedure is chosen
close to zero, one finds the minimum at $\vec \phi_{L}=\vec
\phi_{R}=0$. If one starts at a value beyond the maximum separating
the two minima one obtains the second minimum corresponding to a
superfluid phase. The global minimum of the energy functional
determines the stable quantum phase of the system. The other one
corresponds to a metastable phase.

Thus, energy functionals such as the ones in
Fig.~\ref{fig10:spin1-super-efunctionals} signal metastable phases,
first-order phase transitions and hysteric behavior of the
system; they do not allow the same analysis as the spinless
case. The stability analysis of the $\vec \phi=0$ fixed point does not
answer the question, if there is a second stable fixed point and if it
is energetically lower or higher. To determine the quantum phase we
numerically calculate the energy functional and analyze its local
minima.

Due to the spinor nature of the order parameter additional properties
of the superfluid phases arise.  The spin-dependent interaction
changes the symmetry of the energy functional
$E[\vec\phi_L,\vec\phi_R]$ in the $\phi_0$-$\phi_1$ plane for $\vec \phi_L$
as well as $\vec \phi_R$. For $U_2=0$ the energy functional
is rotationally symmetric around $\phi_1=\phi_0=0$.  For
anti-ferromagnetic interactions $U_2>0$ the bosons form a polar
superfluid, i.e.~the spin-dependent interaction energy is minimized by
$\langle \hat{\textbf{S}}_L\rangle =\langle \hat{\textbf{S}}_R\rangle
= 0$~\cite{ho98}.  There are two different classes of polar order
parameters~\cite{pai08}, one is the transverse polar state
\begin{eqnarray}
\label{transpolar}
\begin{pmatrix}
\phi_1 \\ \phi_0 \\ \phi_{-1}
\end{pmatrix}
=c_1 
\begin{pmatrix}
 1 \\
0 \\
1 \\
\end{pmatrix}
\end{eqnarray}
the other one the longitudinal polar state
\begin{eqnarray}
\label{longpolar}
\begin{pmatrix}
\phi_1 \\ \phi_0 \\ \phi_{-1}
\end{pmatrix}
=c_0 
\begin{pmatrix}
 0 \\
1 \\
0 \\
\end{pmatrix},
\end{eqnarray}
where $c_0$ and $c_1$ are real numbers. Here we omitted the index
labeling the left and right site because the structure of the order
parameter is the same for both of them.

In Fig.~\ref{fig11:spin1-efsurf} we plot the energy functional $E[\vec
  \phi_L, \vec \phi_L]$ as a function of the order parameter for a
symmetric unit cell and a point in parameter space at which there are
two local minima corresponding to a longitudinal and a transverse
polar superfluid phase as well as a local minimum signaling a stable
Mott insulating phase. Since the unit cell is chosen to be symmetric
($\varepsilon=0$) the order parameters in the left and the right site
are the same ($\vec\phi_L=\vec\phi_R$). The two minima at a
non-vanishing order parameter are degenerate, because both correspond
to $\langle \hat{\textbf{S}}_L\rangle =\langle
\hat{\textbf{S}}_R\rangle = 0$ and therefore suffer the same
spin-dependent energy shift of the on-site interaction. Due to the
special form of the two superfluid phases given in
Eq.~(\ref{transpolar}) and Eq.~(\ref{longpolar}) the superfluid minima
are always on the $\phi_1=0$ and the $\phi_0=0$ axes, respectively. 
This justifies why we chose $\phi_0=0$ in
Fig.~\ref{fig10:spin1-super-efunctionals}. The additional minimum in
Fig.~\ref{fig11:spin1-efsurf} at $\phi_0=\phi_1=\phi_{-1}=0$ is the
global minimum and corresponds to the Mott-insulating phase; the two
degenerate minima corresponding to a non-vanishing order parameter
belong therefore to two degenerate metastable phases.

In the ferromagnetic case ($U_2 <0$) there is only one superfluid
order parameter which is given by
$(\phi_1,\phi_0,\phi_{-1})^T=c\ (1,\sqrt{2},1)^T$, where $c$ is a real
number. Again, we suppressed the index labeling the left and right
site because it is the same for both. In the ferromagnetic case, the
spin-dependent interaction in Eq.~(\ref{spin-mfham}) has the same sign
as the tunneling term and therefore does not create metastable quantum
phases.

\subsection{The phase diagram}
\label{sec:spin1phasediagram}

\begin{figure}[tb]
\begin{center}
\includegraphics[width=0.5\textwidth]{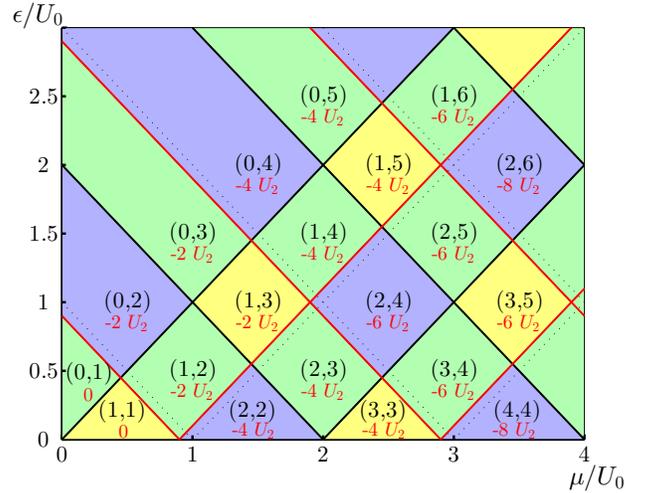}
\caption{ (Color online) Phase boundaries of spin-1 bosons in a
  two-dimensional superlattice for $t_i=t_e=0$, described by the
  Hamiltonian (\ref{spin-mfham}). The red lines denote deviations
  in the phase boundaries relative to the spinless case (see
  Fig.~\ref{fig4:spinless-mottphases}), black solid lines to phase
  boundaries which are not changed and black dotted lines to shifted
  phase boundaries of the spinless case. 
  Each Mott diamond is labeled by its atom number configuration
  $(n_L,n_R)$ (in black) and the energy penalty due to 
  spin-dependent interactions (in red below).  
  The green diamonds correspond to odd-even particle
  number configurations, the yellow ones to odd-odd, and the blue ones
  to even-even configurations.}
\label{fig12:mottphases}
\end{center}
\end{figure} 

In this section we calculate the phase diagram of spin-1 atoms in
superlattices. We focus on the differences to the spinless case which
was discussed in Sec.~\ref{sec:spinlessphasediagram}.

\begin{figure}[tb]
\begin{center}
\includegraphics[width=0.49\textwidth]{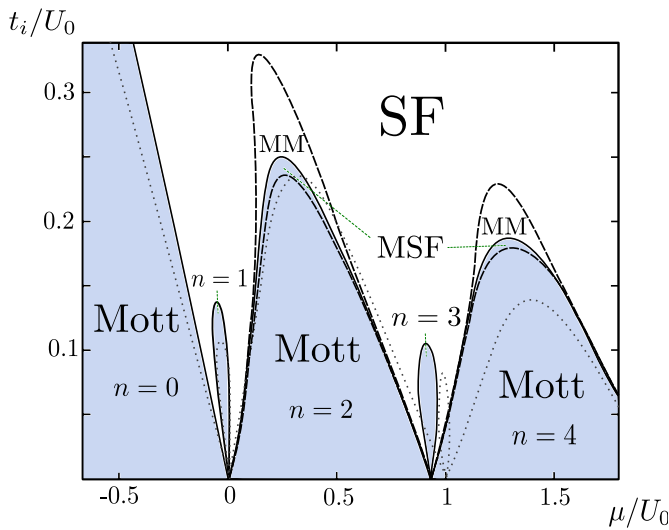}
\caption{ (Color online) Phase diagram for spin-1 atoms in optical
  superlattices ($U_2/U_0=0.04$, $\varepsilon=0$, and $t_i=10
  t_e$). The shaded regions denote Mott-insulating phases. The dashed
  lines are the phase boundaries for metastable phases and the dotted
  lines are the phase boundaries for the spinless case $U_2=0$. The
  regions in which a metastable Mott phase coexists beside the
  superfluid [SF] phase is marked with MM; MSF denotes regions where
  metastable superfluid phases exist alongside the Mott [Mott]
  phase. The Mott lobes are labeled according to the total atom number
  $n$ per double well.}
\label{fig13:ratio10-sww01-cut}
\end{center}
\end{figure} 

For $t_i=t_e=0$ the Hamiltonian~(\ref{spin-mfham}) is diagonal in the
Fock basis and the system supports only Mott phases. Similar to the
spinless case the Mott phases are characterized by $(n_L,n_R)$ and the
boundaries between Mott phases carrying a small number of atoms can be
calculated directly from the collection of eigenstates and
eigenenergies in Table~\ref{table}.

\begin{table}[h]
\begin{tabular}{|c|c||c|c|}
\cline{1-2} \cline{1-2} \cline{3-4}
 state &energy  &state &energy \\
\cline{1-2}\cline{3-4} \hline \hline 
 $E_{00}$ & $0$& $E_{22}$ &  $2U_0 -4 \mu  -4 U_2$ \\ 
\cline{1-2} $E_{01}$ & $-\mu - \varepsilon  $ & $E_{13}$ & $3 U_0-4 \mu - 2 \varepsilon -2 U_2$ \\  
\cline{1-2}$E_{11}$ & $-2 \mu $ &$E_{04} $ & $6 U_0 -4 \mu - 4 \varepsilon -4 U_2$ \\
\cline{3-4} $E_{02}$ & $U_0-2 \mu - 2 \varepsilon - 2 U_2 $ &  $E_{23}$ & $4 U_0-5 \mu  -\varepsilon -4 U_2 $ \\   
\cline{1-2}$E_{12}$ & $U_0-3 \mu -  \varepsilon -2 U_2$ &$E_{14}$ &$6 U_0-5 \mu -3\varepsilon -4 U_2 $ \\
$E_{03}$ & $3 U_0-3 \mu - 3 \varepsilon -2 U_2$  &$E_{05}$ &$10 U_0 -5 \mu- 5\varepsilon  -4 U_2 $\\  
\cline{1-2} \cline{3-4}
\end{tabular}
\caption{Diagonal elements of the Hamiltonian (\ref{spin-mfham}) for
  $t_i=t_e=0$ in Fock space (i.e., $|n_L,n_R\rangle$). Because we are
  interested in ground-state properties for anti-ferromagnetic
  interactions we choose for each atom number configuration the
  smallest spin configuration. For ferromagnetic interactions the highest spin
  configuration is energetically favorable and the table is thus
  changed.}
\label{table}
\end{table}

In Fig.~\ref{fig12:mottphases} we show how the phase boundaries are
shifted compared to the spinless case for anti-ferromagnetic
interactions.  Again, there is a diamond-shaped structure as in
Fig.~\ref{fig4:spinless-mottphases}. The Mott diamonds of
Fig.~\ref{fig4:spinless-mottphases} increase or shrink depending on
their atom number configuration. For anti-ferromagnetic spin
interactions, the strength of the on-site interaction depends upon the
parity of the atom number at each lattice site. An even number of
spin-1 atoms allows the formation of a spin singlet, i.e.~vanishing
total spin per site, which minimizes the on-site repulsion. Odd atom
numbers are penalized, because the spin-singlet wavefunction is
anti-symmetric for an odd atom number and thereby ruled out by
symmetry constraints. Diamonds corresponding to an even particle
number in the left as well as the right well are favored and diamonds
corresponding to an odd-odd configuration are penalized.

The boundary between two Mott diamonds is shifted only if the
spin-dependent energy penalty (or bonus) is different for them
(compare Table \ref{table}).  The value of this shift depends linearly
on $U_2$. In the anti-ferromagnetic case, a phase boundary is either
unshifted or shifted by a constant amount. This is because at each
phase boundary the atom configuration changes only by one atom and the
only possible ground-state spin configurations at each lattice site
are spin singlets ($\langle \hat{\textbf{S}}^2\rangle=0$) and total
spin equals one ($\langle \hat{\textbf{S}}^2\rangle=2$).  This is the
reason why the straight lines in Fig.~\ref{fig4:spinless-mottphases}
are preserved in case of anti-ferromagnetic interactions. Thus, to
determine the shift of the phase boundaries it is enough to examine an
example, say the phase boundary between the Mott diamonds containing
one atom in the left well and the ones containing two atoms in the
left well. The phase boundary follows for positive $\varepsilon$ the
path along $\varepsilon = - \mu + U_0 -2 U_2$, which can be seen by
setting $E_{11}=E_{12}$ or $E_{10}=E_{20}$. The shift of the phase
boundary is therefore $\Delta=\sqrt{2} U_2$.  The diamonds
corresponding to an odd number of atoms in the double well (green
diamonds in Fig.~\ref{fig12:mottphases}) change their size from $1/2$
(in units of $U_0^2$) to $1/2-\Delta^2/U_0^2$.  Double wells carrying
an even number of atoms allow even-even configurations (blue diamonds in
Fig.~\ref{fig12:mottphases} with area $(1/\sqrt{2}+\Delta/U_0)^2$) 
and odd-odd configurations (yellow diamonds in
Fig.~\ref{fig12:mottphases} with area $(1/\sqrt{2}-\Delta/U_0)^2$).

\begin{figure}[tb]
\begin{center} 
\includegraphics[width=0.49\textwidth]{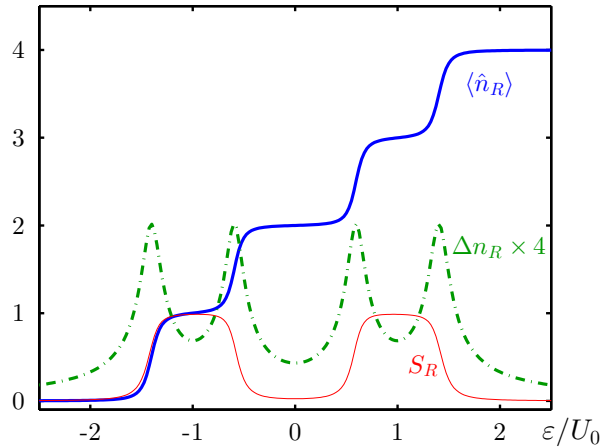}
\caption{(Color online) Bosonic staircase for spin-1 atoms described
  by the Hamiltonian (\ref{spin-mfham}) along the curve $\mu/U_0=1.4$,
  $t_i/U_0=0.05$, $t_e/U_0=0.005$, $U_2/U_0=0.1$ and $\varepsilon/U_0
  \in [-2.5, 2.5]$. The thick solid line shows the particle number in
  the right well $\langle n_R \rangle$ and the dashed line the standard
  deviation $\Delta n_R=\sqrt{\langle n_R^2 \rangle - \langle n_R \rangle^2}$.
  The thin solid line shows $S_R$, the quantum number of the square of the spin
  in the right well, i.e.~$\langle \hat{\textbf{S}}^2_R \rangle =S_R(S_R+1)$.
  The total particle number $\langle \hat n_L + \hat n_R \rangle = 4$
  over the full shown range of $\varepsilon$.  The standard deviation is
  multiplied by 4.}
\label{fig14:spin1-treppe}
\end{center}
\end{figure}

The full ground-state phase diagram for anti-ferromagnetic
interactions, a symmetric double well and $t_i=10 t_e$ is shown in
Fig.~\ref{fig13:ratio10-sww01-cut}. We choose $U_2/U_0=0.04$
corresponding to $^{23}$Na~\cite{burke98}. Spin-dependent interactions
lead to elongated Mott lobes. In general, the spin configuration is
higher in the superfluid phase than in the Mott phase and this leads
to an energy penalty (see Hamiltonian (\ref{spin-mfham})).  Whenever
there is an even number of atoms in a lattice site this effect is
strongest, because an even number of spin-1 bosons can form spin
singlets (see above). In Fig.~\ref{fig13:ratio10-sww01-cut} the Mott
lobe containing four atoms is therefore significantly enlarged, since
this Mott phase corresponds to two atoms in the left as well as the
right site (the unit cells are chosen to be symmetric, i.e.,
$\varepsilon=0$). The Mott lobe $n=2$ is significantly enlarged for
$\varepsilon/U_0=\pm 1$ (not contained in
Fig.~\ref{fig13:ratio10-sww01-cut}), because the atoms pair up on the
left (or right) site and form spin singlets. The phase transitions
between Mott lobes corresponding to an even number of atoms in the
double well are of second order whereas the others are first-order
phase transitions.  For smaller values of $U_2/U_0$ all phase
transitions become first order at a tricritical point, in contrast to
spin-1 atoms in usual lattices for which the boundary of the Mott phase
with one atom per site is always a second-order phase
transition~\cite{krutitsky05}.

To illustrate the phase diagram and to compare it to the case of
spinless bosons we include a bosonic staircase for spin-1 atoms with
anti-ferromagnetic interactions in Fig.~\ref{fig14:spin1-treppe}. In
this figure the occupation number of the right site of each unit cell
$\langle \hat{n}_R \rangle$ is plotted as a function of the energy
offset $\varepsilon$.  We choose parameters so that the system is
Mott-insulating for $-2.5 \leq \varepsilon/ U_0 \leq 2.5 $.
The comparison with Fig.~\ref{fig7:spinless-treppe1} shows that
anti-ferromagnetic interactions shift the steps and make them unequally wide.
The step corresponding to two atoms in the left site and two atoms in the right
site allows the formation of spin singlets ($\langle
\hat{\textbf{S}^2}_L\rangle =\langle \hat{\textbf{S}}^2_R\rangle = 0$)
in both sites. This is energetically favorable compared to the case of an odd
number of atoms on both sites, which form states with $\langle
\hat{\textbf{S}}^2_L\rangle =\langle \hat{\textbf{S}}^2_R\rangle = 2$ (this is analogous  to the changed size of the diamonds in
Fig.~\ref{fig12:mottphases}).

\subsection{Magnetic fields}
\label{sec:mag}
\begin{figure}[tb]
\begin{center}
\includegraphics[width=0.49\textwidth]{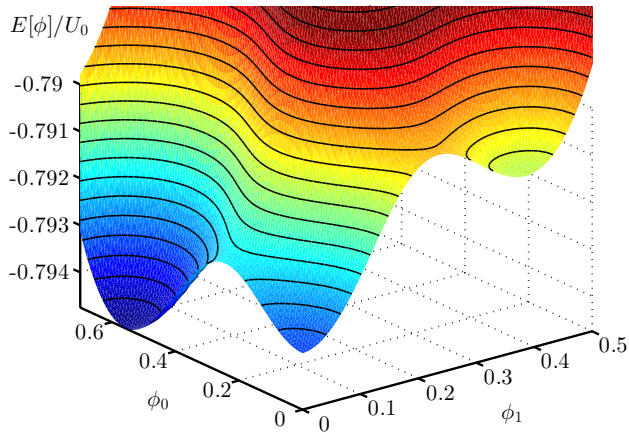}
\caption{ (Color online) Energy functional $E[\vec \phi_L,\vec\phi_R]$
  for a symmetric unit cell ($\varepsilon = 0$) and anti-ferromagnetic
  interactions $U_2/U_0 = 0.04$ in the presence of a magnetic field
  $q/U_0 = 0.002$.  The parameters are chosen identical to
  Fig.~\ref{fig11:spin1-efsurf} ($t_i/U_0=0.3$, $t_e/U_0=0.03$ and
  $\mu/U_0 = 0.25$). Because the unit cell is chosen to be symmetric
  the order parameters of the left and the right site are equal
  ($\vec\phi_L = \vec\phi_R$) and the corresponding indexes are
  suppressed. Because of the rotational symmetry $\phi_1=\phi_{-1}$.
}
\label{fig15:spin1-mag-efsurf}
\end{center}
\end{figure}

\begin{figure}[tb]
\begin{center}
\includegraphics[width=0.49\textwidth]{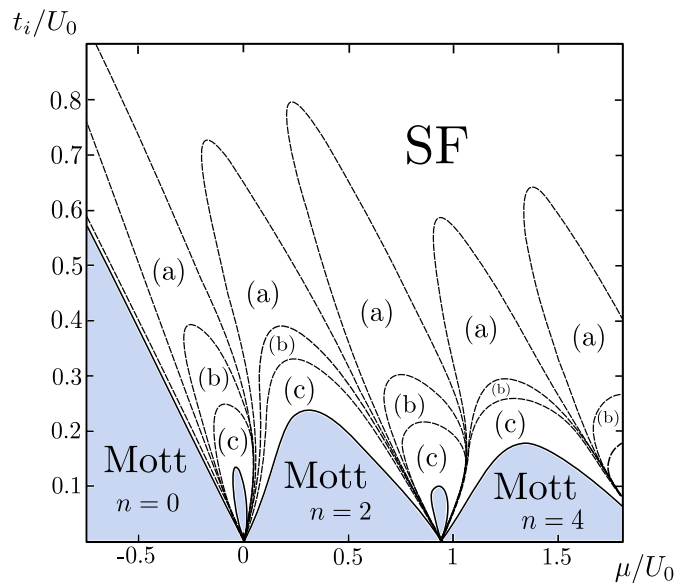}
\caption{ (Color online) Phase diagram for spin-1 atoms in optical
  superlattices with a magnetic field described by the Hamiltonian
  (\ref{hammitmag}) ($U_2/U_0=0.04$, $\varepsilon=0$, $t_i=10 t_e$, and
  $q/U_0=0.02$). The shaded regions denote Mott-insulating phases. The
  dashed lines are the phase boundaries for metastable phases. There
  are no metastable quantum phases in the Mott-insulating region, but
  additional metastable phases in the superfluid one. The quantum
  phase (a) denotes a superfluid region where there are two metastable
  phases. There is a metastable superfluid phase of the transverse
  polar state and energetically higher a metastable Mott phase. In the
  quantum phase (b) there are also both metastable phases but the
  metastable Mott phase of the transverse polar state is energetically
  lower than the metastable superfluid phase. The quantum phase (c)
  contains only one metastable phase of the transverse polar state
  which is a Mott phase.}
 \label{fig16:spin1-supermag}
\end{center}
\end{figure}

Finally, we want to examine the effect of weak magnetic fields on the
phase diagram of spin-1 atoms in optical lattices. Spinor
Bose-Einstein condensates are realized in optical traps, because
strong magnetic fields would break the degeneracy of the different
hyperfine levels and align the atom spins, thus creating a condensate
of effectively spinless bosons. Nevertheless it is interesting to
include weak magnetic fields in the calculations because magnetic
shielding is needed but can be done only to a certain degree.
To incorporate the effect of a magnetic field in the Bose-Hubbard
model for spin-1 bosons one can add a term to the Hamiltonian given in
Eq.~(\ref{spin-mfham}) which includes the Zeeman shift of the energy
levels~\cite{imambekov04}.  The first contribution of a magnetic field
$\vec B= (0,0,B)$ chosen along the $z$-axis (which is our quantization
axis) is a linear Zeeman shift,
\begin{eqnarray}
\hat H_\text{LZ} = p \sum_{i=L,R}  \sum_{\sigma}m_{i \sigma} \hat
n_{i \sigma} =   p \ \hat S^{tot}_z,
\end{eqnarray}
where $p=g \mu_B B$ and $ \hat n_{i \sigma}$ is the particle number
operator for the $i$th site that gives the number of bosons in the
$m$th hyperfine state.  Although the linear Zeeman shift alters the
energy levels considerably, it can in general be neglected. The
interactions within the atomic cloud do not change the overall
magnetization \cite{stenger98,rodriguez11,stamperkurn12} and the
linear Zeeman shift can be gauged away. For a given system
with fixed magnetization the main impact of a magnetic field is given
by the quadratic Zeeman effect,
\begin{eqnarray}
\label{hammitmag}
\hat{H}_\text{QZ} &=& q \sum_{i=L,R}  \sum_{\sigma}m_{i \sigma}^2 \hat n_{i \sigma},
\label{mitspinandmag}
\end{eqnarray}
which is added to the Hamiltonian~(\ref{spin-mfham}).  The magnitude
of the quadratic Zeeman shift is given by $q=q_0 B^2$, where
e.g. $q_0=h \times 390 $ Hz/G$^2$ for sodium~\cite{stenger98}.

The quadratic Zeeman effect affects the phase diagram considerably.
The local minima of the energy functional $E[\vec\phi_L,\vec\phi_R]$ belonging to
transverse and longitudinal polar superfluids are no longer degenerate
(see Fig.~\ref{fig15:spin1-mag-efsurf}). For positive $q$ the
longitudinal superfluid states are always energetically favored, for
negative $q$ the transverse ones. For anti-ferromagnetic spin
interaction new classes of metastable quantum phases arise, see
Fig.~\ref{fig16:spin1-supermag}. It is important to notice, that even
very weak magnetic fields ($q/U_0=0.002$ in
Fig.~\ref{fig15:spin1-mag-efsurf}) change the properties of the ground
state energy functional substantially.
In Fig.~\ref{fig16:spin1-supermag} we choose the same parameters as in
Fig.~\ref{fig11:spin1-efsurf} just adding a very weak magnetic
field. The magnetic field causes a quantum phase transition from the
superfluid phase to a Mott insulating one since the minimum at
$\phi_{1}=\phi_{-1}=0$, $\phi_{0}=0.55$ is now the global
one. Additionally, there are now two metastable quantum phases. The
first is a metastable Mott-insulating phase at
$\phi_0=\phi_1=\phi_{-1}=0$. The second one is a metastable transverse
polar superfluid phase at $\phi_{1}=\phi_{-1}=0.38$ and $\phi_{0}=0$.

Finally, we have also analyzed the ferromagnetic case.  The presence
of a magnetic field changes the energy functional of the ground state
in such a way that first-order phase transitions and metastable phases
are possible.

\section{Discussion and conclusions}

We have analyzed the ground-state phase diagram for spinless and spin-1
atoms in period-2 superlattices. The dynamics within the
unit cells was included exactly and the tunneling between unit cells in a
mean-field approximation. We discussed several methods to treat this
mean-field Hamiltonian and concluded that in the spinless case a
simple stability analysis is sufficient to determine whether the
system is Mott-insulating or superfluid. Using this method, we
have first calculated the phase diagram for spinless bosons in optical
superlattices. In agreement with previous studies~\cite{buonsante05},
we found a contraction of Mott lobes to loops for specific values of
the energy offset. We have presented a detailed study of the various
Mott phases which emerge when the chemical potential and the energy
offset are varied. Furthermore, we have calculated the occupation numbers of single sites
and found single-atom resonances in the Mott-insulating regime.  These
were known for isolated double-well potentials and were generalized in
this paper to the case of superlattices. We found clear fingerprints
of single-atom resonances also in the density of condensed bosons.

In the case of spin-1 atoms the mean-field Hamiltonian shows a much
richer quantum phase diagram. For anti-ferromagnetic interactions all
Mott lobes are elongated towards higher tunneling amplitudes. Mott
lobes with an even number of atoms at each lattice site are especially
favored because their atomic spins can couple to form spin singlets.
A small, non-vanishing order parameter leads to increased atom number
fluctuations and higher spin configurations, and, as a consequence, to higher on-site
repulsion.  Thus, the ground-state energy for small values of the
order parameter is increased.  For certain parameter regimes this
leads to the appearance of two local minima of the ground state energy
functional (one at vanishing order parameter, one at a finite value of
the order parameter) separated by an energy barrier.  The higher one
corresponds to a metastable quantum phase. Thus, the system shows a
hysteretic behavior and the phase transitions are of first order,
whereas they are strictly of  second order for the spinless case. For a
realistic value of the spin-dependent interactions for sodium it
depends on the parity of the atom number if the phase transition of a
specific Mott lobe becomes first order or remains second order. For
smaller values of the spin-dependent interactions all Mott lobes show
first-order phase transitions, contrary to the case of spin-1 atoms in
usual lattices where the phase transition of the Mott lobe with one
atom per site remains second order for all values of the spin
interaction~\cite{krutitsky05}. Because of the richer properties of
the energy functional in the spin-1 case it is no longer possible to
determine the quantum phase of the system with a stability analysis of
the Mott phase only. The ground-state energy functional for each point
in parameter space has to be analyzed, and we have given a detailed analysis
of the size of the various Mott phases in the atomic limit and pointed
out the differences to the case of spinless bosons. We have also compared the single-atom resonances for spin-1 atoms with
the case of spinless atoms and concluded that spin-dependent
interactions change the occupation numbers of individual lattice
sites. Spin-1 atoms in optical superlattices are therefore a model for
mesoscopic magnetism.

Finally, we have discussed the effects of magnetic fields by using an
effective Hamiltonian which includes a quadratic Zeeman shift.  For
anti-ferromagnetic interactions magnetic fields break the degeneracy between
different polar superfluid phases. This leads to new classes of
metastable phases and thus an even richer phase diagram.  In the
ferromagnetic case magnetic fields cause first-order phase transitions
and metastable phases. These results apply to spin-1 atoms in
superlattices as well as in usual lattices.

In conclusion, we have shown that spinor bosons in optical
superlattices show a rich phenomenology of different phases. We hope
that our analysis will be a useful guideline for future experiments.

\section{Acknowledgments}

We thank R.~Fazio for discussions. This paper was financially
supported by the Swiss SNF, the NCCR Nanoscience, and the NCCR Quantum
Science and Technology.

\end{document}